\documentclass[12pt]{article}

\usepackage{amsfonts}
\usepackage[dvips]{graphics}

\begin{document}
\begin{center}{\large \bf Geodesic Flow on the $n$-Dimensional Ellipsoid as a
Liouville Integrable System}
\end{center}
%\vskip .51cm
\begin{center}{Petre Di\c t\u a\footnote{email: dita@hera.theory.nipne.ro}}
\end{center}
%\vskip .21cm
\begin{center}
{National Institute of Physics \& Nuclear Engineering}\\
{Bucharest, PO Box MG6, Romania}
\end{center}
\vskip.31cm
\abstract{We show that the motion on the $n$-dimensional ellipsoid is
complete integrable by exihibiting $n$ integrals in involution. The system is
separable at classical and quantum level, the separation of classical
variables being realized by the inverse of the momentum map. This system 
is a generic one in a new class of $n$-dimensional complete integrable
Hamiltonians defined by an arbitrary function $f(q,p)$ invertible with respect
to momentum $p$ and rational in the coordinate $q$.}
\vskip1.51cm 
Complete integrable Hamiltonian systems in the Liouville sense are a main
subject 
of interest in the last decades both at classical as well as
quantum level. Classical examples of such systems are the geodesic flow on
the triaxial ellipsoid, Neumannn's dynamical system and the integrable cases
of the rigid body motion. The number of interesting examples increased after
the seminal papers by Lax \cite{Lx} and  by Olshanetski and Peremolov
\cite{OP}, especially in connection with the inverse scattering method and
the relation with  classical simple Lie algebras.

 The aim of
this paper is the study of  the geodesic motion on the
$n$-dimensional ellipsoid, which is a direct generalization of the
$2$-dimensional case studied by Jacobi  \cite{Ja}, since it seems that no
solution is known for this system in the case $n>2$\,. We obtain a complete
description of the problem at the classical level by finding $n$  prime
integrals in involution, the separation of variables, the explicit solution of
the Hamilton-Jacobi equation and the equations of geodesics.  We also  show
that the Schr\"odinger equation separates. This system is to a great extent
universal among the various integrable systems and generates a new class of
completely integrable models.

The Lagrangean for a particle of unit mass constrained to move on the
$n$-dimensional ellipsoid   $${x_1^2\over a_1}+{x_2^2\over a_2}+\dots +{x_{n+1}^2\over
a_{n+1}}=1\eqno(1)$$ where $a_i,\, i=1,2\dots, n+1$ are positive numbers
$a_i\in {\mathbb{ R}}_{+}^{n+1}$, is 
$${\mathfrak{ L}}={1\over 2}\sum_{i=1}^{i=n+1}\dot{x}_i^2\eqno(2)$$
The preceding equations define a constrained system and the  model can be
formulated in terms of constrained dynamical variables with Dirac brackets,
or in an unconstrained form with canonical Poisson brackets. In the
following we will use a third way, the classical one, which consists  in 
reducing the Lagrangean by eliminating one degree of freedom \cite{Wh} using
the equation of constraint (1).  
The simplest way to do that  would be to resolve Eq.(1) with respect to the
last coordinate and use it in the free Lagrangean, Eq.(2), but the drawback is
that we obtain a non-diagonal metric. We follow here the Jacobi idea  which
was to find a clever parametrization such that the coresponding metric should
be diagonal \cite{Ja,VK}.

For what follows it is useful to define two polynomials
$$P(x)=\prod_{i=1}^{n+1}(x-a_i)\qquad Q(x)=\prod_{i=1}^{n}(x-u_i)\eqno(3)$$ 
where $a_i$ and $u_i,\, i=1,\dots,n$ are the positive numbers
entering the parametrization of the ellipsoid and  the ellipsoidal
coordinates, respectively. The  orthogonal parametrization of the quadric
(1) is   given by \cite{Kn} $$x_j^2={a_j\,Q(a_j)\over P'(a_j)},\qquad
j=1,2,\dots,n+1\eqno(4)$$ where $P'(a_j)=d\,P(x)/d\,x|_{x=a_j}$, and the
ellipsoidal coordinates $u_1,\dots,u_n$ satisfy
$a_1<u_1<a_2<\dots<u_n<a_{n+1}$. Using this parametrisation  in 
Eq.(2) we find $${\mathfrak{ L}}=-{1\over
8}\sum_{i=1}^{i=n}g_{ii}\dot{u}_i^2\eqno(5)$$ where the (diagonal) metric is
given by $g_{ii}=u_i\,Q'(u_i)/P(u_i)\,,  i=1,2,\dots,n$, and
$Q'(u_i)=d\,Q(x)/d\,x|_{x=u_i}$. Defining as usual the generalized momenta by
$p_i={\partial {\mathfrak{L}}/\partial{\dot{u}_i}}$ and  using  the Legendre
transform  we find the Hamiltonian of the problem $${\mathfrak
{H}}=\sum_{i=1}^{i=n}p_i\,\dot{u}_i- {\mathfrak{
L}}=-2\sum_{i=1}^{i=n}g^{ii}p_i^2\eqno(6)$$ where $g^{ii}={P(u_i)/
u_i\,Q'(u_i)}$. We define now the symmetric functions of the polynomials
$ Q^{(j)}(x)=Q(x)/(x-u_j)$ $${ Q^{(j)}(x)}=\sum_{k=0}^{n-1}x^k\,
S_{n-k-1}^{(j)},\quad j=1,2\dots,n\eqno(7)$$ where $S_{0}^{(j)}=1,\quad
S_{1}^{(j)}=-(u_1+\dots+u_{j-1}+u_{j+1}+\dots+u_n)$, etc.  The upper
index means that the coordinate $u_j$ does not enter the symmetric sum
$S_{k}^{(j)}$, $k=1,\dots,n-1$. We define the following functions
$$H_{k}=\sum_{l=1}^{n}
S_{k-1}^{(l)}g^{ll}p_l^2=\sum_{i=1}^{n}S_{k-1}^{(i)}\,{P(u_i)\over
u_i\,Q'(u_i)}\,p_i^2,\qquad \quad k=1,2\dots,n  \eqno(8)$$ where  $H_1$
differs from ${\mathfrak{{H}}}$ by a numerical factor. 
A careful inspection of the Eqs.(8) shows that for each degree of freedom the
contribution to the Hamiltonian $H_k$ is given by the product of two different
factors. The first one depends on the "Vandermonde" structure
$f_1=S_{k-1}^{(i)}/Q'(u_i)$ and the second one $f_2=(P(u_i)/u_i)p_i^2$ depends
on the "singularities", i.e. the hyperellitic curve defined by the parameters
$a_i$.

Let  $g(p,u)=\cal{H}({\it p,u})$ be  an arbitrary function depending on the
canonical variables $p$ and $u$ which is
invertible with respect to the momentum $p$.  As we will see later
the invertibility condition is necessary for the separation of variables in
the  Hamilton-Jacobi equation. In particular we may suppose that
$\cal{H}({\it p,u})$ is an one-dimensional Hamiltonian. For each
$n\in\mathbb{N}$ we define   an $n$-dimensional  integrable model  by giving
$n$ integrals in involution 
$$\mathbb{H}\,_k(p,u)=\sum_{i=1}^{n}{S_{k-1}^{(i)}\over
Q'(u_i)}g(p_i,u_i),\qquad k=1,2,\dots,n \eqno(8')$$      Our main result is
contained in the following proposition.
%%%%%%%%%%%%%%%%%%%%%%%%%%%%%%%%%%%%%%%%%%%%%%%%%%%%%%%%%%%%%%%%%%%%%% 
\vskip2mm {\bf Proposition}.{\it Let $M^{2n}\simeq T^*(R^n)$ be the canonically
symplectic phase space  of the dynamical system defined by the Hamilton
function Eq.(6). Then\\ i) the functions $H_i,\,\,i=1,2,\dots,n$ are in
involution $$\{H_i,H_j\}=0,\quad i,j=1,2,\dots,n$$ ii)  the momentum map
is given by  $${\cal E}: M^{2n}\rightarrow R^n: M_{\bf h}=\{(u_i,p_i):
H_i=h_i, \quad i=1,2,\dots,n\},\, h_i\in \mathbb{R}$$ then ${\cal
E}^{-1}(M_{\bf h})$ realizes the separation of variables  giving an explicit
factorisation of Liouville's tori into one-dimensional ovals\\ iii) canonical
equations are integrable by quadratures} 
%%%%%%%%%%%%%%%%%%%%%%%%%%%%%%%%%%%%%%%%%%%%%%%%%%%%%%%%%% 
 \vskip2mm In the
following we sketch a proof of the above proposition. As it will be easily
seen the same proof is also true for the Hamiltonians defined  in
Eqs.(8$'$). \\
{\it Proof}.\\ \vskip2mm {\it i)}
We calculate the Poisson bracket $$\{H_k,H_l\}=\sum_{j=1}^n\left({\partial
H_k\over\partial u_j}{\partial H_l\over\partial p_j}-{\partial
H_k\over\partial p_j}{\partial H_l\over\partial u_j}\right)= 2\sum_{j=1}^{n} 
p_j{P(u_j)\over u_jQ'(u_j)}\left(S_{l-1}^{(j)}{\partial H_k\over\partial
u_j}-S_{k-1}^{(j)}{\partial H_l\over\partial u_j}\right)=$$
$$2\sum_{j=1}^n\sum_{i=1}^n p_i^2 p_j{P(u_j)\over u_jQ'(u_j)}
\left[S_{l-1}^{(j)}{\partial\over\partial u_j}\left({S_{k-1}^{(i)} P(u_i)\over
u_i Q'(u_i)}\right)-S_{k-1}^{(j)}{\partial\over\partial
u_j}\left({S_{l-1}^{(i)} P(u_i)\over u_i Q'(u_i)}\right)\right]=$$ $$
\sum_{j=1}^n\sum_{i=1}^n p_i^2 p_j{P(u_j)\over u_jQ'(u_j)}
\left[{\partial\over\partial u_j}\left({S_{l-1}^{(j)}S_{k-1}^{(i)} P(u_i)\over
u_i Q'(u_i)}\right)-{\partial\over\partial
u_j}\left({S_{k-1}^{(j)}S_{l-1}^{(i)} P(u_i)\over u_i
Q'(u_i)}\right)\right]=$$ $$\sum_{j=1}^n\sum_{i=1}^n p_i^2 p_j{P(u_j)\over
u_jQ'(u_j)} \left[{\partial\over\partial
u_j}\left(\left(S_{l-1}^{(j)}S_{k-1}^{(i)}-
S_{k-1}^{(j)}S_{l-1}^{(i)}\right){P(u_i)\over u_i Q'(u_i)}\right)\right]$$ The
last step was possible because the symmetric functions $S_k^{(j)}$ and
$S_l^{(j)}$ depend on all $u_1,u_2,\dots,u_n$, but $u_j$. Looking at the last
expression it is easily seen that the partial derivative with respect to $u_j$
vanishes for $i=j$. For $i\not=j$ we have to show that  
$${\partial\over\partial u_j}{S_{l-1}^{(j)}S_{k-1}^{(i)}-
S_{k-1}^{(j)}S_{l-1}^{(i)}\over{u_i-u_j}}=0$$ but this is a direct consequence
of the following identities $${\partial\over\partial u_j}S_{k-1}^{(i)}= -
S_{k-2}^{(i,j)} \quad {\rm and}\quad
S_{k-1}^{(i)}-S_{k-1}^{(j)}=(u_i-u_j)S_{k-2}^{(i,j)}$$ where the upper index
$(i,j)$ means that the corresponding expression does not depend on both $u_i$
and $u_j$. In this way we have  shown that  $\{H_k,H_l\}=0$.   \vskip2mm {\it
ii)} The surface $ M_{\bf h}=\{(u_i,p_i): H_i=h_i, \quad i=1,2,\dots,n\},\,
h_i\in {\mathbb {R}} $ is given by the  system of equations
$$\sum_{i=1}^{n}S_{k-1}^{(i)}\,{P(u_i)\over u_i\,Q'(u_i)}\,p_i^2=h_k,\qquad
k=1,2\dots,n\eqno(9)$$ and resolving it with respect to $p_i$ is equivalent to
the calculation of the following determinant $$V_n=\left| \begin{array}{ccccc}
1&1&\dots&\dots&1\\ S_1^{(1)}&S_1^{(2)}&\dots&\dots&S_1^{(n)}\\
\dots&\dots&\dots&\dots&\dots\\ \dots&\dots&\dots&\dots&\dots\\
S_{n-1}^{(1)}&S_{n-1}^{(2)}&\dots&\dots&S_{n-1}^{(n)}\\  \end{array}\right|$$
which is equal to the Vandermonde determinant, i.e. $$V_n=\prod_{1\le i<j\le
n}(u_j-u_i)$$
Let $V_{n-1}^{(j)}$ be the determinant obtained by removing the $j$th column
and the last row in $V_n$ and $W_{n,j}$ the determinant obtained by replacing
the $j$th column of $V_n$ by $(h_1,\dots,h_n)^t$. It is easily seen that
$$V_{n-1}^{(j)}= \prod_{1\le k<l\le
n  \atop k\not= j\not= l}(u_l-u_k)$$ i.e. $V_{n-1}^{(j)}$ is the Vandermonde
determinant of the variables $u_1,\dots,u_n$ , but $u_j$. We have the
identities  
$$\prod_{j=1}^n  V_{n-1}^{(j)}=(V_n)^{n-2}$$
$${V_n\over V_{n-1}^{(j)}}=(-1)^{n-j}Q'(u_j),\qquad
j=1,\dots,n$$
$$W_{n,j}=(-1)^{n-j}V_{n-1}^{(j)}\sum_{i=0}^{n-1}h_{n-i}u_j^i, \qquad
j=1,\dots,n$$ Using these identities it is easily seen that ${\cal
E}^{-1}(M_{\bf h})$ is equivalent to  the relations $${P(u_i) p_i^2\over u_i}=
\sum_{k=0}^{n-1}\, h_{n-k}\,u_i^k \qquad i=1,2\dots,n \eqno(10)$$   which
shows that ${\cal E}^{-1}(M_{\bf h})$ is a $n$-dimensional submanifold of
$M^{2\,n}$ and more important $p_i^2$ are functions which depend only on the
variable $u_i$. The last relation shows that the application ${\cal
E}^{-1}(M_{\bf h})$ realizes the separation of variables for the geodesic
motion on the ellipsoid. 

For the Hamiltonians given by Eqs.(8$'$) the preceding equations have the
form
$$g(p_i,u_i)=
\sum_{k=0}^{n-1}\, h_{n-k}\,u_i^k \qquad i=1,2\dots,n \eqno(10')$$
The relations (10-10$'$) have the classical form \cite{Sk}
$$ \varphi(x_i,p_i,h_1,\dots,h_n)=0\,\qquad i=1,2\dots,n$$  which is an
explicit factorization of Liouville's tori into one-dimensional ovals. This is
important for the quantization problem.

With the notation $R(u)=\sum_{k=0}^{n-1}\,
h_{n-k}\,u^k$ the above relations can be written 
$$p_i=\epsilon_i\,\sqrt{u_i\,R(u_i)\over P(u_i)}, \quad i=1,\dots,n$$ 
$$p_i=g^{-1}(R(u_i)), \quad i=1,\dots,n$$ 
where
$\epsilon_i=\pm 1$ and $g^{-1}$ is the inverse of the relation (10$'$) with
respect to the momentum $p$.  
The last relations  allow immediately to resolve the
Hamilton-Jacobi equation because in this case the action splits into a sum of
terms

$$S({\bf h},u_1,\dots,u_n)=S_1({\bf h},u_1)+\dots+ S_n({\bf h},u_n)$$ each of
them satisfying an ordinary differential equation.
 Only   the solution of  the
Hamilton-Jacobi equation for the geodesic motion on the ellipsoid will be
presented the other more general case being similar.
 $$S({\bf h},u_1,\dots,u_n)=
\sum_{i=1}^n\,\epsilon_i\,\int_{u_i^0}^{u_i}\,\sqrt{w\,R(w)\over P(w)}\,dw$$ 

{\it iii)} The above formulae allows us to choose new canonical variables as
follows
${\cal Q}_1=H_1, {\cal Q}_k=H_k$, $k=2,\dots,n$ and the corresponding variables
${\cal P}_i$, $i=1,\dots,n$. The Hamilton equations take the form
$$ \dot{{\cal Q}}_i=   0,\qquad  i=1,\dots,n$$
$$\dot{{\cal P}}_1= -1,\,\, \dot{{\cal P}}_i=0,\,\, i=2,\dots,n$$ and therefore
${\cal Q}_i=h_i,\,\, i=1,\dots,n$ and ${\cal P}_1=-t+g_1,\,\,
{\cal P}_k=g_k,\,\, k=2,\dots,n$, with $g_i,h_i\, \in\,{\mathbb {R}},\,
i=1,\dots,n$.

Because $${\cal P}_i=-{\partial S\over\partial{\cal Q}_i}=-{\partial
S\over\partial h_i}={\epsilon_i\over
2}\int_{u_i^0}^{u_i}{t^{n-i+1/2}\over\sqrt{P(t)R(t)}}\,dt$$ we obtain the
system $$-t\delta_{1j}+b_j={1\over
2}\sum_{i=1}^n\epsilon_i\int_{u_i^0}^{u_i}{t^{n-j+1/2}\over\sqrt{P(t)R(t)}}\,dt,\quad
j=1,\dots,n$$ which gives the implicit equations of the geodesics. In this way
the integration of the Hamilton equations was reduced to quadratures. On the
last  expressions one can see that all the subtleties of the geodesic motion
on the $n$-dimensional ellipsoid are encoded by the hyperelliptic curve
$y^2=P(x)\,R(x)$ whose genus is $g=n$.

For  quantization we use both the forms (9) and (10) and show first that the
quantization of $H_1$  is equivalent to the quantization of Eq.(10). It
is well known that because of the ambiguities concerning the ordering of $p$
and $u$ we must  use the Laplace-Beltrami operator \cite{Po}. Its general form
is $\Delta_n={1\over \sqrt{g}}p_i(\sqrt{g}\,g^{ij}p_j)$, $i,j=1,\dots,n$, where
$g=det(g_{ij})$ and $g_{ij}$ is the metric tensor. Taking into account that
$$g_{ii}=u_i\,Q'(u_i)/P(u_i)=(-1)^{n-i}{V_n\over V_{n-1}^{(i)}}{u_i\over
P(u_i)}$$ after some
simplifications the Schr\"odinger equation generated by the Hamiltonian $H_1$ 
is written in the following form $$-\sum_{i=1}^n{1\over
V_n}{\sqrt{P(u_i)\over u_i}}{\partial\over\partial
u_i}\left((-1)^{n-i}V_{n-1}^{(i)} {\sqrt{P(u_i)\over
u_i}}{\partial\Psi\over\partial u_i}\right)=h_1\,\Psi$$ Since the term
$V_{n-1}^{(i)}$ does not depend on $u_i$ it can be pulled out of the bracket
and the precedent equation takes the form
$$-\sum_{i=1}^n (-1)^{n-i} V_{n-1}^{(i)} {\sqrt{P(u_i)\over
u_i}}{\partial\over\partial u_i}\left( {\sqrt{P(u_i)\over
u_i}}{\partial\Psi\over\partial u_i}\right)=h_1\, V_n\,\Psi$$
   Now  we 
make use of the Jacobi identity for the Vandermonde determinant. With the
above notations the identity  is
 $$\sum_j\,(-1)^{n-j}V_{n-1}^{(j)}\,u_j^k=\delta_{n-1,k}\,V_n$$ 
and using it in the preceding relation we obtain
 $$
\sum_{i=1}^n(-1)^{n-i}V_{n-1}^{(j)}
\left[ {\sqrt{P(u_i)\over u_i}}{\partial\over\partial u_i}\left(
{\sqrt{P(u_i)\over u_i}}{\partial\Psi\over\partial
u_i}\right)+{\sum_{k=0}^{n-1}}\,c_{n-k}\,u_i^k\,\Psi \right]=0$$ 
which is equivalent with $n$ indepedent equations of the form
$${\sqrt{P(u_i)\over u_i}}{\partial\over\partial u_i}\left( {\sqrt{P(u_i)\over
u_i}}{\partial\Psi_i\over\partial
u_i}\right)+\left({\sum_{k=0}^{n-1}}\,c_{n-k}\,u_i^k\right)\,\Psi_i=0\,,\quad
i=1\dots,n\eqno(11)$$  Here $c_1=h_1$ and the other $c_k$ are arbitrary. The
direct approach, starting from  Eq.(10), is simpler the problem being
one-dimensional and one arrives at the same equation, Eq.(11). It has the
advantage that the arbitrary coefficients  $c_k$ are  identified to $c_k=h_k$,
i.e. $c_k$ are the eigenvalues of the Hamiltonians $H_k$.

The Eq.(11) has the general form of a Sturm-Liouville problem 
$$-{d\over dx}\left(p(x){df(x)\over dx}\right)+v(x)f(x)=\lambda\,r(x)f(x)$$
which has to be resolved on an interval $[a,b]$. It is well known that its
eigenfunctions will live in a Hilbert space iff $p(x)\,r(x)>0$ on $[a,b]$. 
If $p(x)$ has a continuous first derivative and $p(x)\,r(x)$  a 
continuous second derivative, then by the following coordinate and function
transforms
 $$\varphi=\int^u \left(r(x)\over p(x)\right)^{1/2}\,dx,\quad
\Phi=(r(u)p(u))^{1/4}\,f(u)\eqno(12)$$ 
the preceding equation takes the standard form
$$-{d^2\Phi\over d\varphi^2}+q(\varphi)\Phi=\lambda\Phi$$
where 
$$q(\varphi)={\mu^{''}(\varphi)\over\mu(\varphi)}-{v(u)\over r(u)},\quad
\mu(\varphi)=(p(u)\,r(u))^{1/4}$$
and $u=u(\varphi)$ is the solution of the Jacobi inverse problem (12).

In our case, Eq.(11), the transformation is
$$\varphi=\int_{u_0}^u \left(u\,R(u)\over P(u)\right)^{1/2}\,du$$
and the Schr\"odinger equation has the  form
$$-{d^2\Phi\over
d\varphi^2}+{\mu^{''}(\varphi)\over\mu(\varphi)}\Phi=h_1\Phi\eqno(13)$$
where $\mu(\varphi)=(R(u( \varphi)))^{1/4}$ and in $R(u)$ we made the rescaling
$h_k\rightarrow h_k/h_1,\,\, k=1,\dots,n$. Thus we have obtained that the
solving of the Schr\"odinger equation (11) is equivalent to the solving of
the motion of one-dimensional   particle in a potential generated by
$R(u(\varphi))$.

For $n=1$ Eq.(13) is the equation of the one-dimensional rotator
$${d^2\Psi\over d\varphi^2}+l^2\Psi=0$$ with the solution
$\Psi(\varphi)={1\over\sqrt{2\pi}}\,e^{i\,l\,\varphi}$, $l\in\mathbb{ Z}$,
etc. In all the other cases we have to make use of the theory of hyperelliptic
curves, $\theta$-functions  and/or hyperelliptic Abelian functions in order to
obtain explicit solutions. This problem will be treated elsewhere.
What is remarkable is that the solving of the classical problem, or the solving
of the associated  Schr\"odinger equation leads to the use of the same
mathematical formalism, $\theta$-functions or hyperelliptic Kleinian functions.

However there is a simpler alternative to Eq.(13), namely by the change of
variable $$\varphi=\int^u\,\sqrt{t\over P(t)}dt$$ the Eq.(11) takes the form
$${d^2\Psi\over d\varphi^2}+
\left(\sum_{k=0}^{n-1}h_{n-k}\,u(\varphi)^k\right)\Psi=0$$ which is the
equation of a  particle moving in a potential generated by the integrals in
involution.

As concerns the quantization of the Hamiltonian $\mathbb{H}\,_1$ it depends on
its explicit form and we do not pursue it here.

From the proof of our results it follows that the hyperelliptic curve was
only a tool in obtaining  the separation of variables Eq.(10); in fact the
separation was a direct consequence of the properties of the
Vandermonde determinant. 

In the following we exibit a few examples of new $n$-dimensional integrable
models. Two models which show that  the dimension $n$ of the system has no
direct connection with the number of zeros and/or poles of the function
$g(p,u)$ could be:  $g(p,u)=(sin \,u/ u)\,p^2$ and  $g(p,u)=tg\,
u\,e^{\,\alpha \,p}$, the first example being a function with a denumerable
number of zeros and the second one with a denumerable number of poles and
zeros, in both cases the hyperelliptic curve being of infinite genus. Other
examples are deduced for example from the many-body elliptic Calogero-Moser
\cite{CM}, or the elliptic Ruijenaars models \cite{Rj}. Starting 
with the corresponding one-dimensional Hamiltonians $H_{CM}(p,u)= f(p,u)=p^2/2
+\nu^2{\wp}_{\tau}(u) $ and  $f_R(p,u)= cosh(\alpha\,p)\sqrt{1-2\,(\alpha\nu)^2
{\wp}_{\tau}( u)}$ respectively, where ${\wp}_{\tau}(u)$ is the Weierstrass function, we
obtain $n$-dimensional models. 

In conclusion   we discovered in this paper a new class of complete integrable
systems which allows to uncover the origin of their integrability or
solvability property.  We have shown that there is a simple and general
mechanism allowing us to construct complete integrable Hamiltonian systems
with an arbitrary number of degree of freedom, and for all these systems the
separation of classical variables   is given by the inverse of the
momentum map.

\end{document}